\documentclass[11pt]{article} 
\usepackage{hyperref} %\pdfoutput=1 
\usepackage{graphicx}

\begin{document} 
 
\title{Antibubbles in a cyclone eyewall} 
 
\author{D. Terwagne, G. Delon, N. Vandewalle, H. Caps, and S. Dorbolo
\\\vspace{6pt} GRASP, Opto-fluidique\\
D\'epartement de Physique B5 \\
Universit\'e de Li\`ege \\
B-4000 Li\`ege \\
Belgium } 
 
\maketitle

\begin{abstract} 
A negative bubble, coined antibubble, is composed by a thin air shell that is immersed in a 
soapy mixture.  A large vortex is generated in the liquid using a mixer.  An antibubble is 
then created close to the surface.  The antibubble is fastly attracted by the vortex.  It 
rotates around the core and comes closer and closer.  When the stress is large enough, the 
vortex deforms the antibubble that winds around the eye vortex.  The antibubble looks like 
a spiral.  Under some conditions,  the antibubble splits into several antibubbles that are 
ejected out of the eye vortex while the largest part is still trapped the vortex.  This latter is 
elongated and is absorbed to the bottom of the tank before popping. \\
This abstract is related to a fluid dynamics video for the gallery of fluid motion 2009.
\end{abstract} 
 
% main text 
An antibubble is the negative object of a bubble.  Instead of having a thin liquid film that separates air and air, an antibubble is a thin air film that separates liquid and liquid.  Both liquids are of the same composition and must be a water mixture with a surfactant.  This object has been discovered in 1934 by Hughes and Hughes \cite{hughes} and revisited by Stong 40 years after \cite{stong}.  Antibubbles are obtained by pouring a soapy mixture into a tank that contains the same mixture.  The flow rate and the size of the impacting jet are the relevant parameters \cite{stone}.  An antibubble is not stable compared to a bubble because no force is opposed to the thinning of the air film.  Their lifetime is only governed by the slow drainage of the air from the bottom to the top of the antibubble \cite{epl}.  Antibubbles are very fragile.  Any object that is able to adsorb on the air layer can provoke the breakdown of the antibubble \cite{njp}.

In the presented videos, we propose to introduce antibubbles into a whirlpool in order to test their properties under a shearing stress.  Roughly speaking, the velocity gradients are vertical in the eye of the whirl and horizontal out of the eye wall.  The used mixture is composed by water and Dreft (trademark P\&G).  A whirl is generated by using a magnetic stirrer.  When the whirl is stabilized, antibubbles are generated in the vicinity of the eye (see Fig.1a).  According to the initial position of the antibubble, the antibubble may be trapped around the eye wall (Fig.1b and c) or in the eye of the whirl (Fig.1d).  The deformations are very different.  Out of the whirl eye, the antibubble is horizontally stretched and wound around the eye.  On the other hand, if the antibubble is trapped in the center of the whirl, it is vertically elongated and it goes down the pool.  For large antibubble, it is possible to split the antibubble into two new antibubbles. 

 SD thanks FNRS for Þnancial support  \\
Part of this work has been supported by COST P21: physics of droplets (ESF)
Videos can be found at the following links:
\href{http://ecommons.library.cornell.edu/bitstream/1813/13863/3/TERWAGNE_mpeg1.mpg}{VIDEO 1}
\href{http://ecommons.library.cornell.edu/bitstream/1813/13863/2/TERWAGNE_mpeg2.mpg}{VIDEO 2}

\begin{figure*}
\begin{center}
\includegraphics[width=10cm]{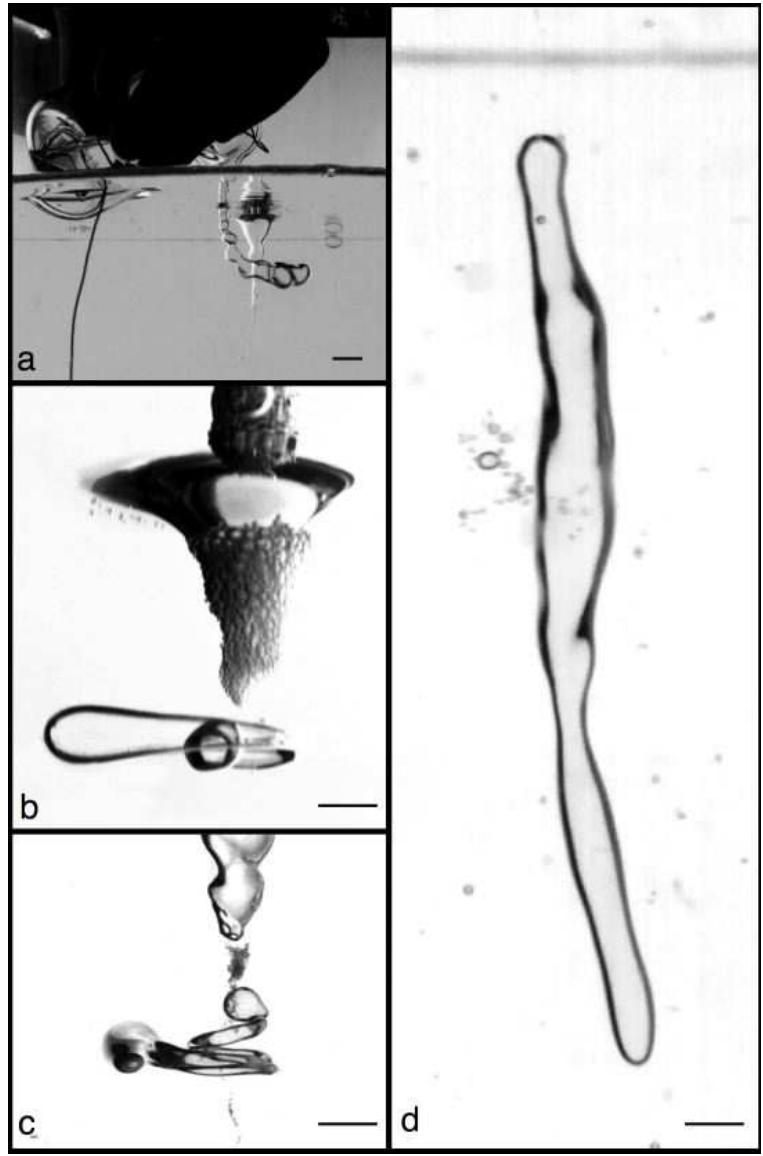}

\end{center}
\caption{(a) Creation of an antibubble close to the whirlpool. (b) An antibubble is trapped by the whirl and is stretched around the eye wall. (c) An antibubble is wound around the whirl. (d) An antibubble is located in the eye.  It is vertically stretched.  The small lines in the different figures represent one centimeter.}
\end{figure*}

\end{document}